\documentclass{IEEE_lsens}
\usepackage{textcomp}
\usepackage{graphicx}
\usepackage{orcidlink}
\usepackage[nolist, nohyperlinks]{acronym}
\usepackage[noadjust]{cite}
\usepackage{multirow}
\usepackage{multicol}
\usepackage[T1]{fontenc} 
\usepackage{amsmath}
\usepackage[cmintegrals]{newtxmath}
\usepackage{bm}
\usepackage{array}
\usepackage{adjustbox}
\usepackage{url}

\providecommand{\hypersetup}[1]{\relax}

\hypersetup{pdftitle={Low-complexity Attention-based Unsupervised Anomalous Sound Detection exploiting Separable Convolutions and Angular Loss},
pdfsubject={Audio Anomaly Detection},
pdfauthor={Michael Neri},
pdfkeywords={Unsupervised Anomaly Detection, Audio Processing, Deep Learning, Wavegram, Attention, Explainability}}
\begin{acronym}
\acro{AE}{Autoencoder}
\acro{AI}{Artificial Intelligence}
\acro{ASD}{Anomalous Sound Detection}
\acro{AUC}{Area Under the Curve}
\acro{pAUC}{partial Area Under the Curve}
\acro{BCE}{Binary Cross-Entropy}
\acro{CNN}{Convolutional Neural Network}
\acro{DNN}{Deep Neural Network}
\acro{DoS}{Denial-of-Service}
\acro{ELU}{Exponential Linear Unit}
\acro{FCN}{Fully Convolutional Network}
\acro{FPR}{False Positive Rate}
\acro{GNN}{Graph Neural Network}
\acro{ICA}{Independent Component Analysis}
\acro{IR}{Information Retrieval}
\acro{IM}{Inlier Modeling}
\acro{MSA}{Multivariate Statistical Analysis}
\acro{RNN}{Recurrent Neural Network}
\acro{SOTA}{State of the Art}
\acro{STFT}{Short-Time Fourier Transform}
\acro{PCA}{Principal Component Analysis}
\acro{PFA}{Probability of False Alarm}
\acro{USAD}{Unsupervised Sound Anomaly Detection}
\acro{VAE}{Variational Autoencoder}
\end{acronym}

\begin{document}

\markboth{Vol.~1, No.~3, May~2023}{0000000}


\title{Low-complexity Attention-based Unsupervised Anomalous Sound Detection exploiting Separable Convolutions and Angular Loss}

\author{\IEEEauthorblockN{Michael~Neri~\orcidlink{0000-0002-6212-9139}\IEEEauthorrefmark{1}\IEEEauthorieeemembermark{1}, and Marco~Carli~\orcidlink{0000-0002-7489-3767}\IEEEauthorrefmark{1}}
\IEEEauthorblockA{\IEEEauthorrefmark{1}Department of Industrial, Electronic, and Mechanical Engineering, Roma Tre University, Rome, Italy.\\
\IEEEauthorieeemembermark{1}Graduate Student Member, IEEE\\
\IEEEauthorieeemembermark{2}Senior Member, IEEE}%
\thanks{Corresponding author: M. Neri (e-mail: \href{mailto:michael.neri@uniroma3.it}{michael.neri@uniroma3.it}).}
\thanks{Associate Editor: XXXX XXXXX.}%
\thanks{Digital Object Identifier 10.1109/LSENS.XXXX.XXXXXX}}

%

\IEEELSENSmanuscriptreceived{Manuscript received XXXX XX, XXXX}

\IEEEtitleabstractindextext{%
\begin{abstract}
In this work, a novel deep neural network, designed to enhance the efficiency and effectiveness of unsupervised sound anomaly detection, is presented. The proposed model exploits an attention module and separable convolutions to identify salient time-frequency patterns in audio data to discriminate between normal and anomalous sounds with reduced computational complexity. The approach is validated through extensive experiments using the Task 2 dataset of the DCASE 2020 challenge. Results demonstrate superior performance in terms of anomaly detection accuracy while having fewer parameters than state-of-the-art methods. Implementation details, code, and pre-trained models are available in \href{https://github.com/michaelneri/unsupervised-audio-anomaly-detection}{https://github.com/michaelneri/unsupervised-audio-anomaly-detection}.
\end{abstract}

}

\maketitle

\section{Introduction}

\IEEEPARstart{I}{}n the context of unsupervised anomaly detection, an \textit{anomaly} refers to data patterns that deviate from the expected \textit{normal} behavior~\cite{chandola2009anomaly}. Likewise, \ac{ASD} is the task of understanding whether a sound is \textit{normal} or not (\textit{anomalous})~\cite{Chen_2023_ICASSP}. \ac{ASD} is applied in the field of machine condition monitoring~\cite{Dohi2022, Dohi2022-2}, medical diagnosis~\cite{Dissanayake_2021_JBHI}, safety and security in urban environments~\cite{Neri_access_2022}, and multimedia forensics~\cite{Valenzise_2007_AVSS, Neri_2022_EUVIP}. Generally, \acp{DNN} are employed for \ac{ASD} due to their ability to identify subtle and unknown anomalous data patterns~\cite{Wilkinghoff_2024_TASL}. Unsupervised or semi-supervised models are generally adopted in \ac{ASD} problems because of the limited availability of anomalous sounds. \ac{SOTA} \ac{USAD} approaches can be classified into two categories~\cite{Wu_2023_AppliedAcoustics, Guan_2023_ICASSP}: \textit{reconstruction-based} and \textit{classification-based}. In the first scenario, models are based on the hypothesis that only non-anomalous samples, which have been analyzed during training, can be effectively retrieved after lossy compression, e.g., \ac{AE}~\cite{ Koizumi_2019_TASL}. In~\cite{Suefusa_2020_ICASSP} a \ac{DNN} has been designed to interpolate masked time bins of the log-Mel spectrogram. Similarly, in~\cite{Dohi_2021_ICASSP} normalizing flows have been used for estimating the probability density of normal data. However, these models suffer from generalization problems, e.g., an anomalous sample may be correctly reconstructed by an \ac{AE}~\cite{Bovenzi_2023_ComputerNetworks}. Classification-based approaches, instead, compute the anomaly score exploiting probability-based distances between prediction and ground truth, e.g., cross-entropy. The classification is carried out on metadata, which can be the identification number of a specific machine that produced the sound. The design rationale is that a model cannot classify successfully the metadata associated with a sound if it is anomalous~\cite{Giri_2020_DCASE, Wu_2023_AppliedAcoustics, Guan_2023_ICASSP}. The use of metadata as an auxiliary loss function allows the modeling of the probability distribution of normal data, namely \ac{IM}~\cite{kawaguchi2021description}. One such model, STgram-MFN~\cite{Liu_ICASSP_2022}, extracts temporal and spectral features to classify the IDs of machines using ArcFace~\cite{Deng_CVPR_2019}. Similarly, in~\cite{Choi_2023_ARXIV} two novel angular losses ArcMix and Noisy-Arcmix have been designed to enhance the compactness of intra-class distribution during the classification of IDs.  Differently, the authors in~\cite{Guan_ICASSP_2024} involved contrastive learning in the pre-training to reduce distances between pairs of feature embeddings from the same machine IDs.   


\begin{figure}[ht!]
    \centering
    \includegraphics[width=1\linewidth]{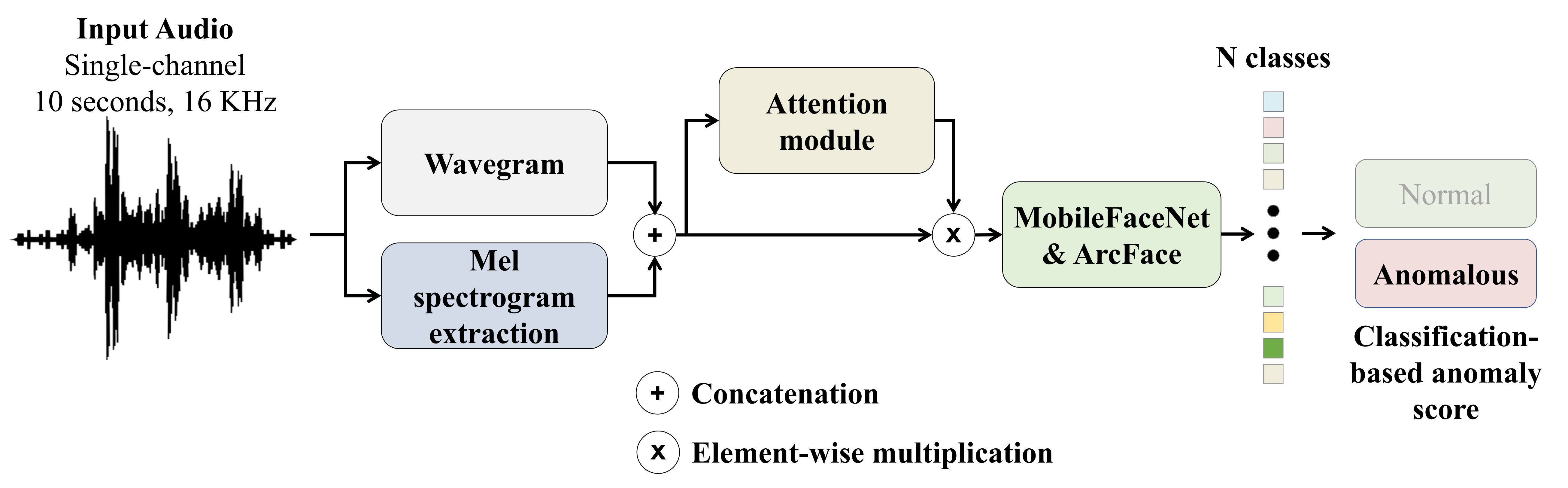}
    \caption{Description of the proposed pipeline for \ac{USAD}.}
    \label{fig:method}
\end{figure}

However, it is important to consider the computational complexity in the context of \ac{USAD}. The response time of an anomaly detector is critical to limit the damage caused by an anomalous event~\cite{Koizumi_2019_TASL}. Hence, this work also analyzes the computational complexity of SOTA approaches in terms of the amount of learnable parameters. Moreover, it is often challenging to interpret why these models flag certain audio segments as anomalies due to their black-box nature. To address this, for the first time in the literature, we employ an attention module~\cite{Neri_TASLP_2024} to provide explanations for the decisions made by the anomaly detection system. The attention mechanism highlights which parts of the input are most influential in the model's anomaly detection, thereby enhancing the interpretability of the model's outputs.

\begin{figure*}[ht!]
    \centering
    \includegraphics[width=0.6\linewidth]{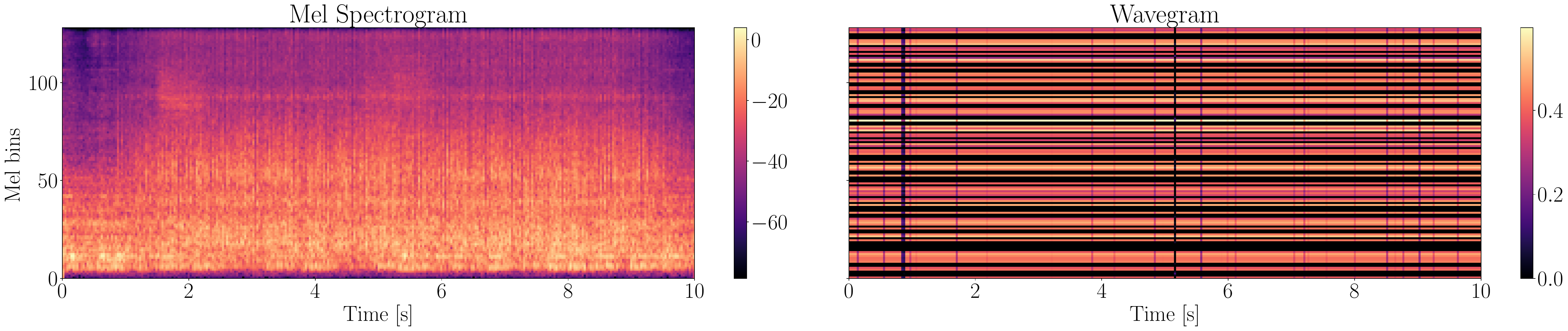}
    \caption{Example of acoustic features $X$ from an anomalous sound of Task 2 DCASE 2020 dataset.}
    \label{fig:spec-heat}
\end{figure*}

\begin{figure*}[ht!]
    \centering
    \includegraphics[width=0.6\linewidth]{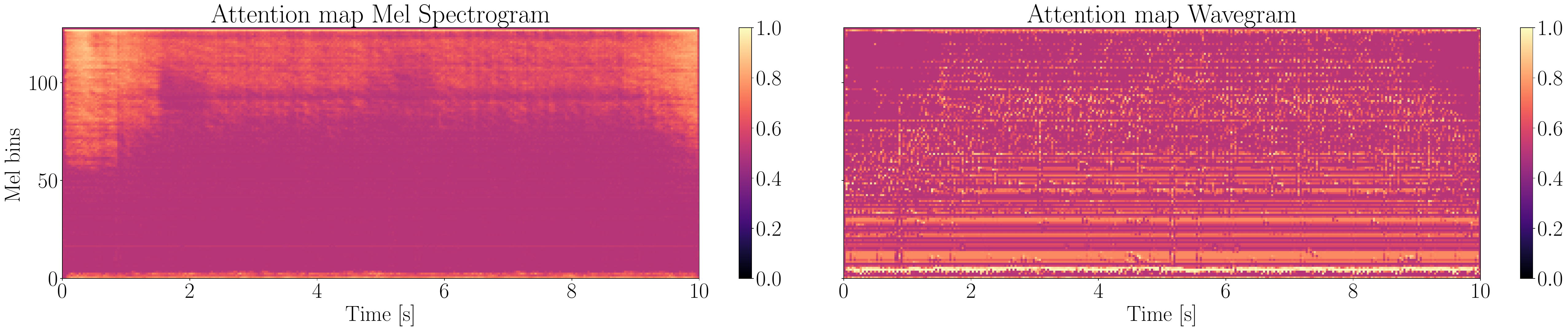}
    \caption{Attention maps $H = f_{\mathrm{ATT}}(X)$ obtained from the attention module with acoustic features in Figure~\ref{fig:spec-heat}.}
    \label{fig:attentions}
\end{figure*}

To summarise, the contributions of this work are:
\begin{itemize}
    \item Definition of an attention module focused on identifying time-frequency anomalous pattern detected both in the log-Mel spectrogram and from the learned representation, i.e., Wavegram~\cite{Kong_2020_TASLP}.
    \item Use of separable convolutions to reduce the computational complexity of the model, decreasing by approximately $13\%$ of the number of learnable parameters concerning the top-tier approaches of the literature;
    \item 
    Statistical analysis of the attention maps highlights the importance of high-frequency bins in the log-Mel spectrogram as the main cue for the identification of anomalous sounds in this scenario. Moreover, a comparison with SOTA approaches, in terms of performance and computational complexity, is carried out.
\end{itemize}

The rest of the paper is organized as follows: Section~\ref{sec:method} details the proposed approach, providing insights on the feature extraction and anomaly score computation. Details of the dataset, metrics, and experimental results are explained in Section~\ref{sec:exp}. Finally, the conclusions are drawn in Section~\ref{sec:conclusion}.

\section{Proposed method}\label{sec:method}
In this section the proposed unsupervised approach for audio anomaly detection is detailed. The goal is to determine whenever a single-channel audio signal with $l$ samples $\mathbf{x} \in \mathbb{R}^{1 \times l}$ is anomalous without using in training the binary anomaly label $y \in \mathbb{Z}^2$. To do so, we employ time-frequency representations as features, namely log-Mel spectrogram and Wavegram, to jointly identify patterns in time and frequency since audio signals are generally non-stationary~\cite{Purwins_JSTSP_2019}. In conjunction with an attention module and angular loss, an efficient \ac{DNN} is proposed for classification-based anomaly detection. The overall architecture is shown in Figure~\ref{fig:method}.

\subsection{Feature extraction}
Initially, a pre-processing stage is employed to extract the complex \ac{STFT} $\mathrm{STFT}\{\mathbf{x}\}$ from the audio signal $\mathbf{x}$. This transform is performed using a Hann window of length $64$~ms with $50\%$ overlap. The selection of the window function is critical since windowing in the time-domain results in a convolution in the frequency domain, disrupting the spectral characteristics of the audio signal. Hann window mitigates this problem thanks to its characteristic of having the localization of spectral energy around the normalized frequency $w = 0$, minimizing spectral leakage~\cite{prabhu2014window}. Length and overlap of windows are consistent with those found in the literature for \ac{ASD}. Next, a log-Mel spectrogram $X_{\mathrm{Mel}} \in \mathbb{R}^{t \times f}$ is extracted using a Mel filterbank $\mathrm{H}_{\mathrm{Mel}}(\cdot)$ as $ X_{\mathrm{Mel}} = 20 \log_{10} \mathrm{H}_{\mathrm{Mel}}(\mathrm{STFT}\{\mathbf{x}\})$, where $t$ and $f$ denote the number of time and frequency bins, respectively. 

In~\cite{Kong_2020_TASLP} Wavegram is introduced as a new learned time-frequency representation for audio tagging. In particular, Wavegram is designed to capture relevant time-frequency cues for the classification that may go unnoticed like hand-crafted log-Mel spectrograms due to its lossy representation~\cite{Kong_2020_TASLP}. Within the scope of \ac{USAD}, several methods have been based on Wavegram by applying a 1D convolution that acts as a learnable \ac{STFT}~\cite{Liu_ICASSP_2022}. Next, the features have been further processed by layer normalization and 1D convolutions with small kernel sizes~\cite{Chen_2023_ICASSP, Choi_2023_ARXIV}. To reduce the computational complexity, in this work, Wavegram consists only of a separable 1D convolutional layer with $f$ strided filters to mimic the windows' overlap in the \ac{STFT} computation. Finally, the log-Mel spectrogram and the output of Wavegram $X_{\mathrm{Wave}} \in \mathbb{R}^{t \times f}$ are concatenated along the channel dimension $X = [ X_{\mathrm{Mel}}, X_{\mathrm{Wave}} ] \in \mathbb{R}^{t \times f \times 2}$. An example of input acoustic features is depicted in Figure~\ref{fig:spec-heat}.

\subsection{Attention module}
The attention module is responsible for learning an attention map $H \in \mathbb{R}^{+t \times f \times 2}$ from the log-Mel spectrogram and the Wavegram. Its objective is to emphasize regions of features that are most informative for the classification task. This module has been extensively analyzed for evaluating the distance between a microphone and a speaker~\cite{Neri_TASLP_2024}. However, its application in \ac{ASD} has not been investigated yet. In this work, it is denoted as the function $\mathrm{f_{\mathrm{ATT}}}:\mathbb{R}^{t \times f \times 2} \rightarrow \mathbb{R}^{{+t \times f \times 2}}$. It comprises $2$ separable convolutional blocks, having $16$ and $64$ $3 \times 3$ filters, respectively. Then, a $1 \times 1$ convolutional layer, that acts as a linear projection to reduce the number of channels, with two filters, followed by a sigmoid activation function for mapping each pixel into a probability, is used to map the features to yield the $t \times f \times 2$ attention map. Finally, the weighted acoustic features $\Tilde{X} \in \mathbb{R}^{t \times f \times 2}$ are obtained by element-wise multiplication ($\otimes$) between the two time-frequency representations and the attention map as $\Tilde{X} = f_{\mathrm{ATT}}(X) \otimes X$.
Examples of attention maps are shown in Figure~\ref{fig:attentions}.

\begin{figure*}[ht!]
    \centering
    \includegraphics[width=0.6\linewidth]{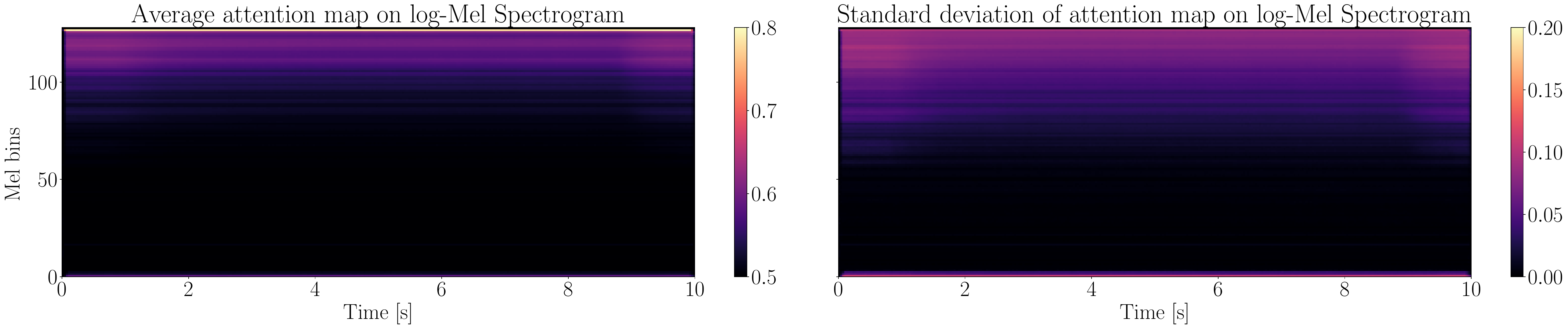}
    \caption{Mean and standard deviation of the attention map of the log-Mel spectrogram on testing set.}
    \label{fig:spec-mel}
\end{figure*}

\begin{table*}[ht!]
\caption{Comparison with SOTA methods. \textbf{Bold} and \underline{underline} are used to highlight first and second-best results, respectively.}
\label{tab:results}
\centering
\adjustbox{max width=0.75\textwidth}{%
\begin{tabular}{c|cc|cc|cc|cc|cc|cc}
\hline \hline 
\multirow{2}{*}{Methods} & \multicolumn{2}{c|}{Fan} & \multicolumn{2}{c|}{Pump} & \multicolumn{2}{c|}{Slider} & \multicolumn{2}{c|}{Valve} & \multicolumn{2}{c|}{ToyCar} & \multicolumn{2}{c}{ToyConveyor}  \\ 
        & AUC [\%] & pAUC [\%] & AUC [\%] & pAUC [\%] & AUC [\%] & pAUC [\%] & AUC [\%] & pAUC [\%] & AUC [\%] & pAUC [\%] & AUC [\%] & pAUC [\%]  \\ \hline
IDNN~\cite{Suefusa_2020_ICASSP} & $67.71$ & $52.90$ & $73.76$ & $61.07$ & $86.45$ & $67.58$ & $84.09$ & $64.94$ & $78.69$ & $69.22$ & $71.07$ & $59.70$ \\
MobileNetV2~\cite{Giri_2020_DCASE} & $80.19$ & $74.40$ & $82.53$ & $76.50$ & $95.27$ & $85.22$ & $88.65$ & $87.98$ & $87.66$ & $85.92$ & $69.71$ & $56.43$ \\
Glow-Aff~\cite{Dohi_2021_ICASSP} & $74.90$ & $65.30$ & $83.40$ & $73.80$ & $94.60$ & $82.80$ & $91.40$ & $75.00$ & $92.20$ & $84.10$ & $71.50$ & $59.00$ \\
GMM + Arcface~\cite{Wu_2023_AppliedAcoustics} & $87.97$ & $80.66$ & $\mathbf{95.63}$ & $\underline{85.74}$ & $99.22$ & $97.55$ & $91.26$ & $84.00$ & $95.28$ & $86.91$ & $69.80$ & $61.21$\\
STgram-MFN~\cite{Liu_ICASSP_2022} & $94.04$ & $88.97$ & $91.94$ & $81.75$ & $\mathbf{99.55}$ & $\mathbf{97.61}$ & $99.64$ & $98.44$ & $94.44$ & $87.68$ & $74.57$ & $63.60$  \\
SW-WaveNet~\cite{Chen_2023_ICASSP} & $\underline{97.53}$ & $\underline{91.54}$ & $87.27$ & $82.68$ & $98.96$ & $94.58$ & $99.01$ & $97.26$ & $95.49$ & $\underline{90.20}$ & $\underline{81.20}$ & $\underline{68.20}$ \\
Noisy-ArcMix~\cite{Choi_2023_ARXIV} & $\mathbf{98.32}$ & $\mathbf{95.34}$ & $\underline{95.44}$ & $\mathbf{85.99}$ & $\underline{99.53}$ & $\underline{97.50}$ & $\underline{99.95}$ & $\underline{99.74}$ & $\underline{96.76}$ & $90.11$ & $77.90$ & $67.15$ \\
\hline
Proposed approach & $95.10$ & $87.25$ & $91.97$ & $80.00$ & $99.24$ & $96.10$ & $\mathbf{99.99}$ & $\mathbf{99.96}$ & $\mathbf{96.99}$ & $\mathbf{90.30}$ & $\mathbf{84.59}$ & $\mathbf{73.55}$  \\
\hline \hline
\end{tabular}
}
\end{table*}

\subsection{Data augmentation}
To improve the robustness of the model, we synthetically augment the dataset using mixup~\cite{Zhang_2018_ICLR} in each batch during the training, defined as $\mathbf{x}^{ij} = \lambda \mathbf{x}^i + (1 - \lambda) \mathbf{x}^j$ and $\mathbf{y}^{ij} = \lambda \mathbf{y}^i + (1 - \lambda) \mathbf{y}^j$, 
where $(\mathbf{x}, \mathbf{y})$ is the tuple describing the waveform $\mathbf{x}$ and the one-hot encoded metadata $\mathbf{y} = [y_1, y_2, \ldots, y_c]$ with $c$ classes of a single audio recording under analysis, respectively. $i,j \in \{ 0, 1, \ldots, n-1 \}$ are randomly selected indexes of training audio samples in the batch with size $n$, and $\lambda \sim \mathrm{Beta}(\alpha, \alpha)$ is the mixup coefficient. This augmentation can be performed at different levels of the deep learning architecture, e.g., input level or at intermediate feature levels~\cite{Zhang_2018_ICLR}. In this work, the augmentation procedure is applied to input signals before the preprocessing step, following~\cite{Choi_2023_ARXIV}.

\subsection{Self-supervised anomaly score}
To distinguish between anomalous and normal sound, an anomaly score $\mathcal{A}_{\theta}$ is computed from the predicted metadata and the ground truth. As a classification-based approach if a sound is misclassified, then it is anomalous since the model is trained to correctly classify normal sounds. We utilize ArcFace~\cite{Deng_CVPR_2019} as the classification layer,

\begin{equation}\label{eq:ArcFace}
    \mathcal{L}(\boldsymbol{\theta}, \mathbf{y})_{\mathrm{AF}} = -\mathbf{y}^T \frac{e^{s\cos{\boldsymbol{\theta} + m\mathbf{y}}}}{\sum_{i = 1}^{c} e^{s\cos{\theta_{i}+m\hat{y}_i}}},
\end{equation}

where the angular vector $\boldsymbol{\theta} = [\theta_1, \theta_2, \ldots, \theta_c]$ is obtained for each class by computing $\theta_i = \arccos (\mathbf{w}_i^T\mathbf{h})$, which is the result of the mapping between the features obtained from the classifier $\mathbf{h} \in \mathbb{R}^{h \times 1}$ and learned ArcFace weights $\mathbf{w}_i \in \mathbb{R}^{h \times 1}$ for the $i$-th class. The scalars $s \in \mathbb{R}^+$ and $m \in \mathbb{R}^+$ are the scale and margin coefficients for the ArcFace loss, respectively. As introduced in~\cite{Choi_2023_ARXIV}, the employed loss function for training the model is
\begin{equation}
    \mathcal{L}(\boldsymbol{\theta}, \mathbf{y}, \mathbf{y}^{ij}) = \lambda \mathcal{L}(\boldsymbol{\theta}, \mathbf{y})_{\mathrm{AF}} + (1 - \lambda) \mathcal{L}(\boldsymbol{\theta}, \mathbf{y}^{ij})_{\mathrm{AF}}.
\end{equation}
During the testing phase, as the augmentation is not performed, the anomaly score is computed as $ \mathcal{A}_{\theta} (y, \hat{y}) = \mathcal{L}(\boldsymbol{\theta}, \mathbf{y})_{\mathrm{AF}}$.

\section{Experimental Results}\label{sec:exp}
The Task 2 development dataset of the DCASE 2020 challenge~\cite{Dohi2022} is used to assess the performance of the proposed approach. It encompasses six machines (Fan, Pump, Slider, Valve, ToyCar, and ToyConveyor) and each machine is labeled with a unique identifier to differentiate audio recordings from various machines within the same category. A total of $41$ machines with $10$ seconds of audio signals are collected. 
To assess the performance of the proposed approach, we evaluate the \ac{AUC} and \ac{pAUC} metrics. The latter is the \ac{AUC} over a low \ac{FPR} in the range $[0, p]$ with $p = 0.1$, following~\cite{Koizumi_2020_DCASE}. Our approach is trained to classify the $c = 41$ labels derived from machine types and IDs~\cite{Dohi2022, Dohi2022-2}. For the loss and mixup computation, parameters are set as $\alpha=0.2$, $m=0.7$, and $s=40$, following the guidelines provided by their corresponding works~\cite{Deng_CVPR_2019, Zhang_2018_ICLR}. Log-Mel spectrogram and Wavegram output have $t = 313$ and $f = 128$ bins. The classifier is MobileFaceNet~\cite{Chen_2018_CCBR} which yields a feature vector with dimensionality $h = 128$. The network is optimized using AdamW with a learning rate of $0.0001$, epochs of $300$, and a batch size of $64$. Hyperparameters of the training procedure have been assigned by means of a grid search optimization procedure.


\subsection{Results}
The performance of the proposed approach compared with those obtained with SOTA architectures are represented in Table~\ref{tab:results}. Overall, the proposed approach shows the best performance in three of the six equipment types (Valve, ToyCar, and ToyConveyor). Specifically, the approach achieves SOTA performance on ToyConveyor, which is the most difficult machine in this dataset. This is possible thanks to the combination of Wavegram and log-Mel spectrogram, providing additional cues to the classifier. In the other classes, the performance is still competitive. Generally, Table~\ref{tab:results} can be used as a reference for the selection of the approach that is most suitable to the specific use case. Regarding the computational complexity, Table~\ref{tab:resultsCompl} highlights the number of parameters and the performance of our approach compared with those of the SOTA. Our system offers a good trade-off between model complexity and performance.   

\begin{table}[ht!]
\caption{Number of parameters, average \ac{AUC}, and average \ac{pAUC} of SOTA approaches and proposed method.}
\label{tab:resultsCompl}
\centering
\adjustbox{max width=0.3\textwidth}{%
    \begin{tabular}{cccc}
    \hline \hline 
    Methods & Parameters & \ac{AUC} [\%] & \ac{pAUC} [\%]  \\ 
    \hline
    IDNN~\cite{Suefusa_2020_ICASSP} & $\mathbf{46}$ \textbf{k} & $76.96$ & $62.57$ \\
    MobileNetV2~\cite{Giri_2020_DCASE} & $1.1$ M & $84.34$ & $77.74$ \\
    Glow-Aff~\cite{Dohi_2021_ICASSP} & $30$ M & $85.20$ & $73.90$  \\
    GMM + Arcface~\cite{Wu_2023_AppliedAcoustics} & $1$ M & $89.86$ & $82.68$   \\
    STgram-MFN~\cite{Liu_ICASSP_2022} & $1.1$ M & $92.36$ & $86.34$  \\
    SW-WaveNet~\cite{Chen_2023_ICASSP} & $27$ M & $93.25$ & $\underline{87.41}$  \\
    Noisy-ArcMix~\cite{Choi_2023_ARXIV} & $1.1$ M & $\mathbf{94.65}$ & $\mathbf{89.31}$  \\
    \hline
    Proposed approach & $\underline{884}$ k & $\underline{93.44}$ & $85.71$  \\
    \hline \hline
    \end{tabular}
    }
\end{table}


Table~\ref{tab:hype} shows the selection of parameters regarding the type of features and the dimensionality of the ArcFace layer. The use of Wavegram representation  $[X_{\mathrm{Wav}}]$ in conjunction with the log-Mel spectrogram can improve the performance of the proposed model by $1.17\%$ in terms of \ac{AUC}, albeit being ineffective using it alone. Moreover, the best performance is obtained by setting the dimensionality of the classification layer to $h = 128$.

\begin{table}[ht!]
\caption{Selection of parameters of the proposed approach.}
\label{tab:hype}
\centering
\adjustbox{max width=0.25\textwidth}{%
    \begin{tabular}{cccc}
    \hline \hline 
    Features & $h$  & \ac{AUC} [\%] & \ac{pAUC} [\%]  \\ 
    \hline
    \multicolumn{4}{c}{\textbf{Feature study}} \\
    $[X_{\mathrm{Mel}}]$ & $128$ & $92.26$ & $84.55$\\
    $[X_{\mathrm{Wav}}]$ & $128$ & $63.48$ & $54.12$ \\
    \multicolumn{4}{c}{\textbf{Dimensionality study}} \\
    $[X_{\mathrm{Mel}},X_{\mathrm{Wav}}]$ & $256$ & $90.87$ & $83.94$ \\
    $[X_{\mathrm{Mel}},X_{\mathrm{Wav}}]$ & $64$ & $91.94$ & $85.00$\\
    \hline 
    $[X_{\mathrm{Mel}},X_{\mathrm{Wav}}]$ & $128$  & $\mathbf{93.43}$ & $\mathbf{85.71}$ \\
    \hline \hline
    \end{tabular}
    }
\end{table}

To better explain which parts of the log-Mel spectrogram are relevant for the ID classification, Figure~\ref{fig:spec-mel} shows the average and standard deviation maps on the testing set of the proposed heatmap. Interestingly, the most important frequency bins for the identification of anomalies, i.e., ID misclassification, are contained in the range $[1.7, 8]$ kHz of the log-Mel spectrogram. In addition, the range $[0, 71]$ Hz is also relevant. The rest of the log-Mel spectrogram $[0.071, 1.7]$ kHZ is assigned a value of $0.5$ with zero variance, denoting this region as less important for the ID classification and, thus, less reliable for the identification of anomalies. To assess the impact of both separable convolutions and the attention module, an ablation study has been carried out. Table~\ref{tab:abl_separable} demonstrates the effectiveness of using the attention map in combination with separable convolutions.

\begin{table}[ht!]
\caption{Ablation study.}
\label{tab:abl_separable}
\centering
\adjustbox{max width=0.4\textwidth}{%
    \begin{tabular}{ccccc}
    \hline \hline 
    Methods & Parameters & \ac{AUC} [\%] & \ac{pAUC} [\%]  \\ 
    \hline
    w/o separable convs., w/o $f_{\mathrm{ATT}}$ & $1$ M & $90.50$ & $83.62$ \\ 
    w/o separable convs. & $1$ M & $92.25$ & $84.82$  \\
    w/o $f_{\mathrm{ATT}}$ & $882$ k & $91.72$ & $84.52$ \\
    \hline 
    Proposed approach & $884$ k & $\mathbf{93.43}$ & $\mathbf{85.71}$  \\
    \hline \hline
    \end{tabular}
    }
\end{table}

\section{Conclusions}\label{sec:conclusion}
In this work, a learning-based low-complexity approach is proposed to detect anomalous sound in a machine monitoring scenario. To this aim, a \ac{DNN} is proposed. It exploits an attention module to highlight the most salient time-frequency patterns for identifying machine IDs. Then, an anomaly score is computed from the classification errors between predicted and ground truth metadata. Experimental results demonstrate the validity of the proposed low-complexity model. Although retraining the entire architecture is necessary to handle domain shifts in new environments or with different machines, the approach's low complexity enables efficient fine-tuning with new normal data. Future work will focus on the improvement of the attention module, coping with more complex tasks in the realm of sound anomaly detection such as few and one-shot unsupervised anomaly detection.

\normalsize

\bibliographystyle{IEEEbib}
\bibliography{bibs}

\end{document}